\pdfoutput=1
\documentclass[a4paper]{article}

\usepackage{INTERSPEECH2021}

\usepackage[ruled,vlined]{algorithm2e}
\usepackage{color,cite}
\usepackage{subfigure}
\usepackage{multirow}
\usepackage{hyperref}

\title{Scaling sparsemax based channel selection for speech recognition with\\ ad-hoc microphone arrays}
\name{Junqi Chen, Xiao-Lei Zhang}
\address{CIAIC, School of Marine Science and Technology, Northwestern Polytechnical University, China}
\email{jqchen@mail.nwpu.edu.cn, xiaolei.zhang@nwpu.edu.cn}

\begin{document}

\maketitle

\begin{abstract}
Recently, speech recognition with ad-hoc microphone arrays has received much attention. It is known that channel selection is an important problem of ad-hoc microphone arrays, however, this topic seems far from explored in speech recognition yet, particularly with a large-scale ad-hoc microphone array. To address this problem, we propose a \textit{Scaling Sparsemax} algorithm for the channel selection problem of the speech recognition with large-scale ad-hoc microphone arrays. Specifically, we first replace the conventional Softmax operator in the stream attention mechanism of a multichannel end-to-end speech recognition system with Sparsemax, which conducts channel selection by forcing the channel weights of noisy channels to zero. Because Sparsemax punishes the weights of many channels to zero harshly, we propose Scaling Sparsemax which punishes the channels mildly by setting the weights of very noisy channels to zero only.
Experimental results with ad-hoc microphone arrays of over 30 channels under the conformer speech recognition architecture show that the proposed Scaling Sparsemax yields a word error rate of over $30$\% lower than Softmax on simulation data sets, and over $20$\% lower on semi-real data sets, in test scenarios with both matched and mismatched channel numbers.
\end{abstract}
\noindent\textbf{Index Terms}: distant speech recognition, ad-hoc microphone arrays, channel selection, attention, scaling sparsemax
\section{Introduction}
Distant speech recognition is a challenging problem \cite{haeb2020far}. Microphone array based multichannel speech recognition is an important way to improve the performance \cite{heymann2016neural,xiao2016deep,sainath2017multichannel}. However, because speech quality degrades significantly when the distance between the speaker and microphone array enlarges, the performance of automatic speech recognition (ASR) is upper-bounded physically no matter how many microphones are added to the array \cite{heymann2018performance}. An ad-hoc microphone array is a solution to the above difficulty \cite{raykar2004position}. It consists of a set of microphone nodes randomly placed in an acoustic environment, where each node contains a single-channel microphone or a microphone array.
It can significantly reduce the probability of the occurrence of far-field environments by grouping the channels around the speaker automatically into a local array \cite{zhang2018deep} via \textit{channel reweighting and selection}. Existing channel selection criteria for the ASR with ad-hoc microphone arrays can be divided into two kinds: (i) signal-level-based criteria, such as signal-to-noise-ratio (SNR), and (ii) recognition-level-based criteria, such as word-error-rate (WER).

The first kind of channel selection methods conducts channel selection according to the estimated speech quality of the channels \cite{cossalter2011ad,wolf2014channel,boeddeker2018front,watanabe2020chime}, such as SNR, distance, orientation, envelope variance and room impulse response, by independent estimators from speech recognition systems. After channel selection,
%{the single channel for ASR is picked from the one-best channel directly or fused from N-best channels by beamforming. }
{they fuse the selected channels into a single channel by adaptive beamforming, or pick the one-best channel directly for ASR}.
Although the speech quality based metrics have a strong relationship with the ASR performance in most cases, optimizing the speech quality do not yield the optimal ASR performance.

The second kind aims to conduct channel selection and channel fusion for optimizing the ASR performance directly   \cite{cossalter2011ad,wolf2014channel,li2019multi,li2020practical}. Early methods \cite{cossalter2011ad,wolf2014channel} chose the channels with the highest likelihood of the output after the decoding of ASR.
% Its advantage is boosted in the encoder-decoder structure with attention mechanisms.
Because encoder-decoder structures with attention mechanisms are the new frontier of ASR, the channel selection task has been conducted in the ASR system. \cite{li2019multi} designed a multi-channel encoder structure with a hierarchical attention mechanism, where the output of each channel is first aligned with the first-level attention of the hierarchical attention, followed by the second-level attention called \textit{stream attention} to reweight and fuse the output of all channels. \cite{li2020practical} further improved the hierarchical attention by a two-stage method, which makes all channels share the same encoder in the first stage and then fine-tunes the stream attention in the second stage. It is generalizable to any number of channels. However, the above methods only consider {the} channel reweighting problem with {few} ad-hoc microphone nodes, e.g. no more than 10 microphone nodes, leaving the channel selection problem unexplored. When the environment is large and complicated, and when the number of nodes becomes large as well, it may not be good to take all channels into consideration given that some channels may be too noisy to be helpful.

% (for example, \cite{li2019multi} has two, \cite{watanabe2020chime} has six).
To address the aforementioned problem, this paper proposes two channel selection methods in a conformer-based ASR system for optimizing the ASR performance directly. The contribution of the paper is as follows:
\begin{itemize}
\item The core idea is to replace the Softmax operator in the {stream attention} with two new operators, named \textit{Sparsemax} and \textit{Scaling Sparsemax} respectively, which can force the channel weights of the noisy channels that do not contribute to the performance improvement to zero.

\item Besides, we propose a stream attention based conformer \cite{gulati2020conformer} ASR system with ad-hoc arrays, which is beyond the bidirectional long short-term memory system \cite{li2020practical}

\item At last, different from \cite{li2020practical} which takes all channels into the training of the shared single-channel ASR in the first stage, we first train a single-channel ASR with clean speech data, and then train the Sparsemax and Scaling Sparsemax based stream attention with multi-channel noisy speech data. This training strategy is motivated by the following two phenomena: Given a large ad-hoc microphone array, (i) when we take some very noisy channels into training, the ASR system may not be trained successfully; (ii) the data of all channels is very large.
\end{itemize}
Experimental results with ad-hoc microphone arrays of as many as 30 nodes demonstrate the effectiveness of the proposed methods in both simulated and semi-real data.

%The rest of this paper is organized as follows. Section \ref{sec: system} gives a specific introduction to our system. Section \ref{sec: exp} introduces the data set and experimental setup. Section \ref{sec: result} evaluates our model based on real and simulated data sets. Section \ref{sec: conclu} summarizes our work.

\section{Conformer-based ASR with ad-hoc microphone arrays} \label{sec: system}

 Fig. \ref{fig:singleandmultimodel} shows the architecture of the proposed single-channel and multichannel conformer-based ASR systems, where we omitted residual connections and position embedding modules for clarity. The single-channel system is the first-stage training of the ASR system with ad-hoc arrays, while the multichannel system is the second-stage training.

\subsection{Single channel conformer-based ASR system} \label{sec: single}
Fig. \ref{fig:single} shows the single-channel system. It is trained with clean speech. Specifically, given the input acoustic feature of an utterance $\bm{X} \in \mathbb{R}^{T \times D_x}$
% $\bm{X} = \left\{\bm{x}_t \in \mathbb{R}^{d_x}\ |\ t = 1, \cdots, T \right\}$
 and its target output $\bm{O} \in \mathbb{R}^{L \times D_v}$, where $T$ and $D_x$ is the length and dimension of $\bm{X}$ respectively, and $D_v$ is the vocabulary size.
First, $\bm{X}$ is processed by a convolutional downsampling layer which results in $\bm{\tilde{X}} \in \mathbb{R}^{\tilde{T} \times D_x}$. Then, $\bm{\tilde{X}}$ passes through an encoder $\rm{Enc(\cdot)}$ and a decoder ${\rm Dec}(\cdot)$:
\begin{align}
\bm{H} &= {\rm Enc}(\bm{\tilde{X}}) \\
\bm{c}_l &= {\rm Dec}(\bm{H}, \bm{y}_{1:l-1})
\end{align}
which produces a context vector $\bm{c}_l \in \mathbb{R}^{D_h}$ at each decoding time step $l$ given the decoding output of the previous time steps $\bm{y}_{1:l-1} \in \mathbb{R}^{l-1 \times D_v}$, where $\bm{H} \in \mathbb{R}^{\tilde{T} \times D_h}$ is a high level representation extracted from the encoder.
Finally, $\bm{c}_l$ is transformed to an output vector $\bm{y}_{l}$ by a linear transform. The objective function of the conformer is to maximize:
\begin{equation}
\mathcal{L} = \sum_{l=1}^{L} \log (\bm{y}^T_l \bm{o}_l)
\end{equation}
where $\bm{o}_l$ is the $l$-th time step of $\bm{O}$.

Multi-head attention (MHA) mechanism is used in both the encoder and the decoder, which is the key difference between our conformer architecture and the model in \cite{li2020practical}. Each scaled dot-product attention head is defined as:
\begin{equation}
{\rm Attention}(\bm{Q_i}, \bm{K_i}, \bm{V_i}) = {\rm Softmax}\left(\frac{\bm{Q_i K_i}^T}{\sqrt{D_k}}\right)\bm{V_i} \label{eq:att}
\end{equation}
where $\bm{Q_i} \in \mathbb{R}^{T_1 \times D_k}, \bm{K_i}, \bm{V_i} \in \mathbb{R}^{T_2 \times D_k}$ are called the query matrix, key matrix and value matrix, respectively, $n$ is the number of the heads, $D_k = D_h / n$ is the dimension of the feature vector for each head. Then, MHA is defined as:
\begin{equation}
{\rm MHA}(\bm{Q}, \bm{K}, \bm{V}) = {\rm Concat}(\bm{U}_1, \cdots, \bm{U}_n)\bm{W}^O
\end{equation}
where ${\rm Concat}(\cdot)$ is the concatenation operator of matrices,
\begin{equation}
\bm{U}_i = {\rm Attention}(\bm{Q}_i, \bm{K}_i, \bm{V}_i)
\end{equation}
and $\bm{W}^O \in \mathbb{R}^{D_h \times D_h}$ is a learnable projection matrix.

\subsection{Multichannel conformer-based ASR system} \label{sec: multi}
\begin{figure}[t]	
	\centering
	\subfigure[]{
	\begin{minipage}[t]{0.45\linewidth}
		\centering
		\includegraphics[scale=0.36]{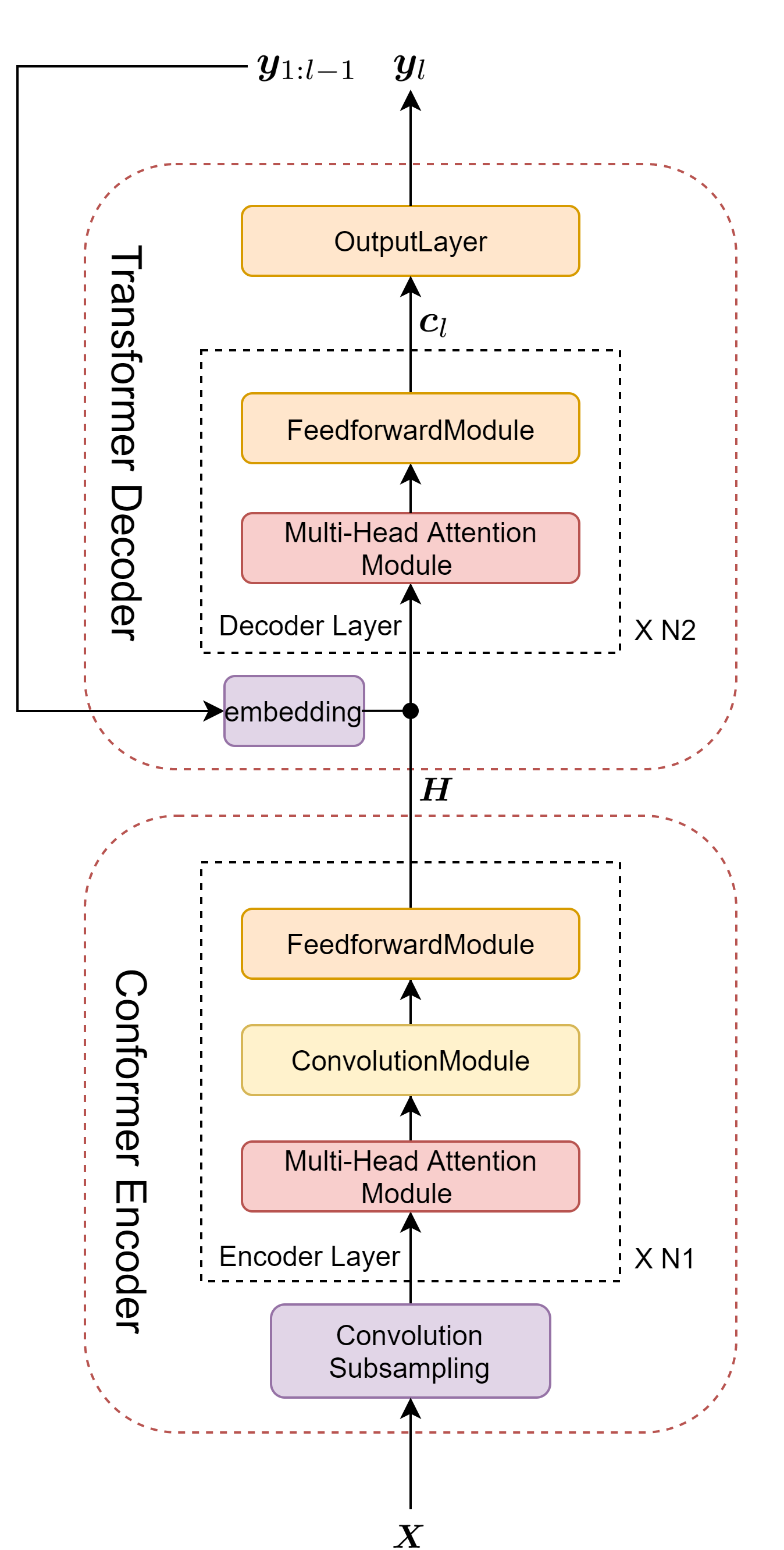}
		\label{fig:single}
	\end{minipage}%
	}
	\subfigure[]{
	\begin{minipage}[t]{0.45\linewidth}
		\centering
		\includegraphics[scale=0.36]{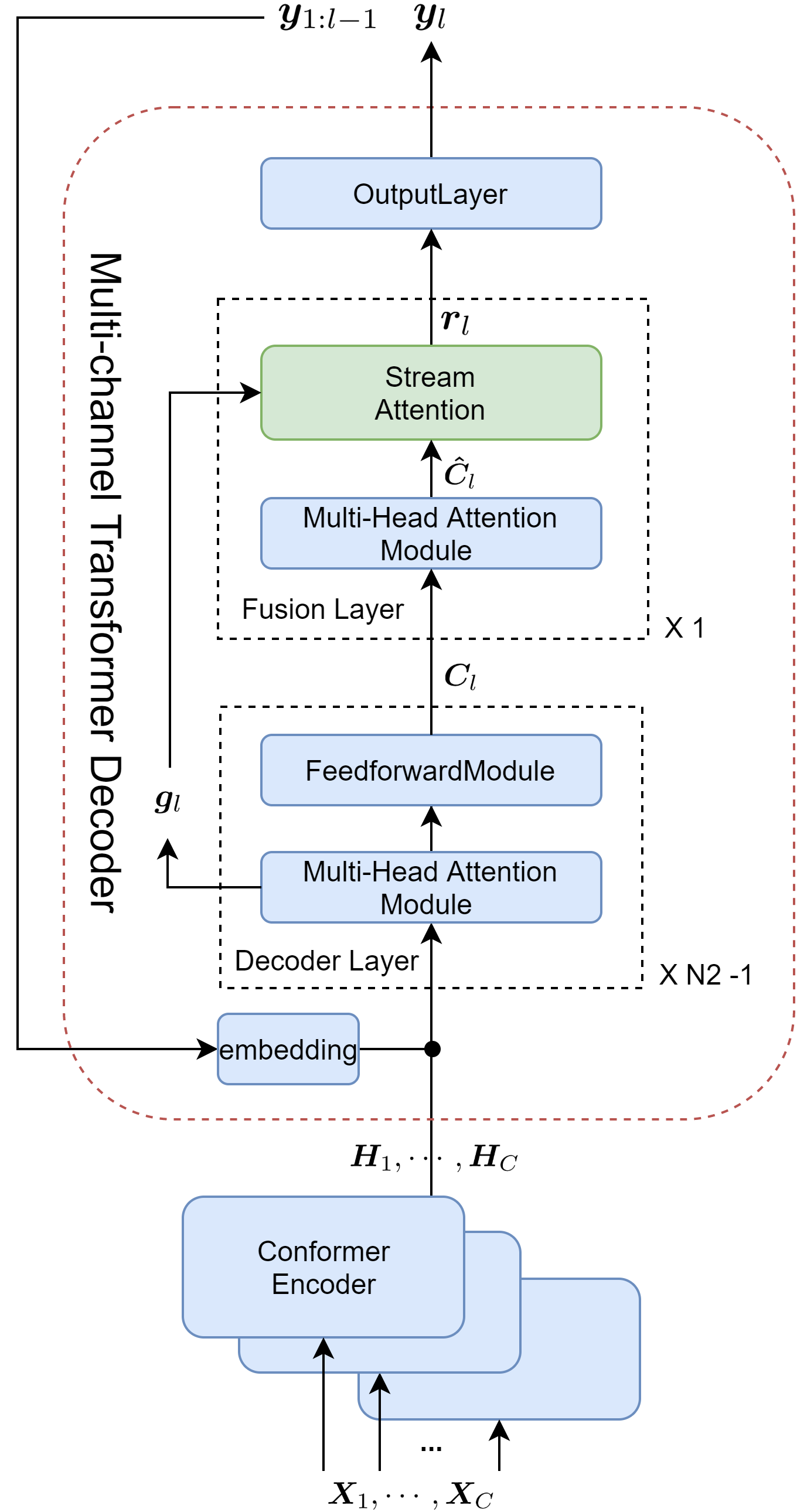}
		\label{fig:multi}
	\end{minipage}%
	}

	\caption{Conformer-based ASR systems. (a) Single-channel model. (b) Multichannel model.}
	\label{fig:singleandmultimodel}
\end{figure}

Fig. \ref{fig:multi} shows the multichannel system. In the figure, (i) the modules marked in blue are pre-trained by the single-channel ASR system, and shared by all channels with their parameters fixed during the training of the multichannel ASR system; (ii) the module marked in green is the stream attention, which is trained in the second stage with the noisy data from all channels.

The architecture of the multichannel system is described as follows.
Given the input acoustic feature of an utterance from the $k$-th channel
$\bm{X}_k \in \mathbb{R}^{T \times D_x},\ k = 1, \cdots, C$ where $C$ represents the total number of channels, we extract high level representations $\bm{H}_k$ from each channels:
\begin{equation}
\bm{H}_k = {\rm Enc} (\bm{\tilde{X}}_k), \ k=1, \cdots, C
\end{equation}
Then, we concatenate the context vectors of all channels:
\begin{equation}
\bm{C}_l = {\rm Concat}\left(\bm{c}_{l,1}, \cdots, \bm{c}_{l,C}\right)
\end{equation}
where
\begin{equation}
\bm{c}_{l,k} = {\rm Dec}(\bm{H}_k, \bm{y}_{1:l-1})
\end{equation}

At the same time, we extract a \textit{guide vector} $\bm{g}_l \in \mathbb{R}^{D_h}$ from the output of the decoder at all previous time steps by:
\begin{equation}\label{eq:xx}
\bm{g}_l = {\rm MHA} (\bm{y}^T_{l-1} \bm{W}^{Y_1}, \bm{y}_{1:l-1} \bm{W}^{Y_2}, \bm{y}_{1:l-1} \bm{W}^{Y_3})
\end{equation}
{where $\bm{W}^{Y_1}, \bm{W}^{Y_2}, \bm{W}^{Y_3} \in \mathbb{R}^{D_v \times D_h}$ denote learnable projection matrices.} The guide vector $\bm{g}_l \in \mathbb{R}^{D_h}$ is used as the input of the stream attention which will be introduced in Section \ref{sec: stream att}.

\section{Variants of stream attention} \label{sec: stream att}

 This section first describes the stream attention framework, and then present the proposed Sparsemax and Scaling Sparsemax respectively.

\subsection{Description of stream attention}
As shown in the fusion layer in Fig. \ref{fig:multi}, the stream attention takes the output of a MHA layer as its input. The MHA extracts high-level context vector by:
\begin{equation}
\bm{\hat{c}}_{l,k} = {\rm MHA} (\bm{c}^T_{l,k} \bm{W}^C, \bm{H}_{k} \bm{W}^{H_1}, \bm{H}_{k} \bm{W}^{H_2})
\end{equation}
where $\bm{W}^C, \bm{W}^{H_1}, \bm{W}^{H_2} \in \mathbb{R}^{D_h \times D_h}$ denote learnable projection matrices.

Then, the stream attention calculates
\begin{equation}
	\bm{r}_l = {\rm StreamAttention} (\bm{Q}, \bm{K}, \bm{V}) \label{eq:rl}
	\end{equation}
with $\bm{Q} = \bm{g}^T_l \bm{W}^G$, $\bm{K}=\bm{\hat{C}}_l \bm{W}^{\hat{C}_1}$ and $\bm{V} = \bm{\hat{C}}_l \bm{W}^{\hat{C}_2}$,
%\begin{equation}
%\bm{r}_l = {\rm StreamAttention} (\bm{g}^T_l \bm{W}^G, \bm{\hat{C}}_l \bm{W}^{\hat{C}_1}, \bm{\hat{C}}_l \bm{W}^{\hat{C}_2}) \label{eq:rl}
%\end{equation}
where $\bm{g}_l$ is the guide vector defined in \eqref{eq:xx},
\begin{equation}
\bm{\hat{C}}_l = {\rm Concat}\left(\bm{\hat{c}}_{l,1}, \cdots, \bm{\hat{c}}_{l,C}\right),  \notag
\end{equation}
and $\bm{W}^G, \bm{W}^{\hat{C}_1}, \bm{W}^{\hat{C}_2} \in \mathbb{R}^{D_h \times D_h}$ are learnable projection matrices. Finally,
we get the output vector $\bm{y}_l$ of the decoder through an output layer from $\bm{r}_l$.

\begin{figure}[t]
	\centering
	\subfigure[]{
	\begin{minipage}[t]{0.31\linewidth}
		\centering
		\includegraphics[scale=0.5]{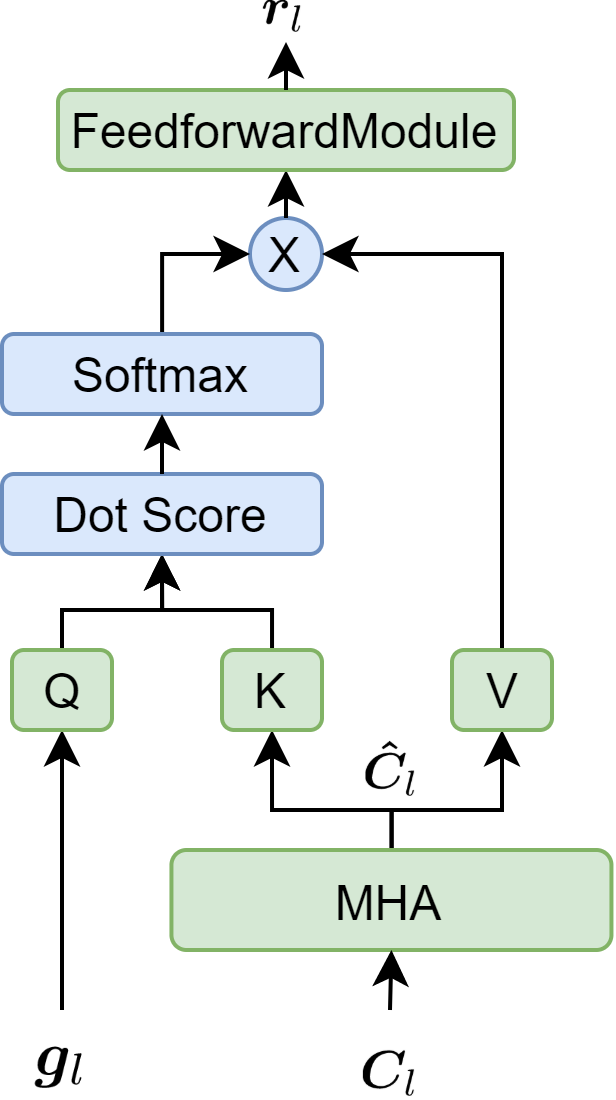}
		\label{fig:softmax}
	\end{minipage}%
	}
	\subfigure[]{
	\begin{minipage}[t]{0.31\linewidth}
		\centering
		\includegraphics[scale=0.5]{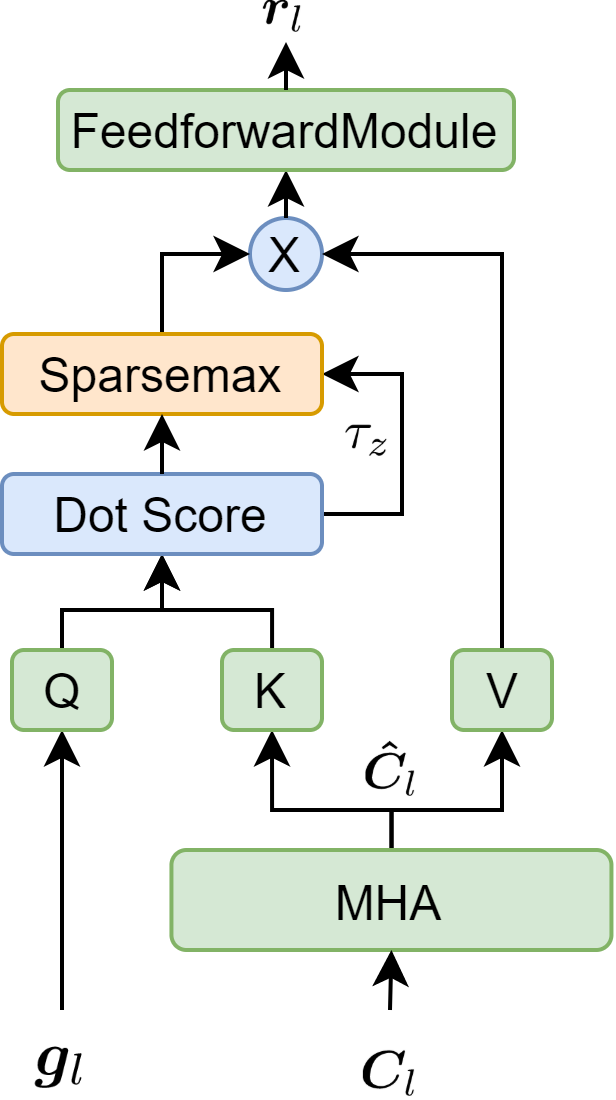}
		\label{fig:sparsemax}
	\end{minipage}%	
	}
	\subfigure[]{
	\begin{minipage}[t]{0.31\linewidth}
		\centering
		\includegraphics[scale=0.5]{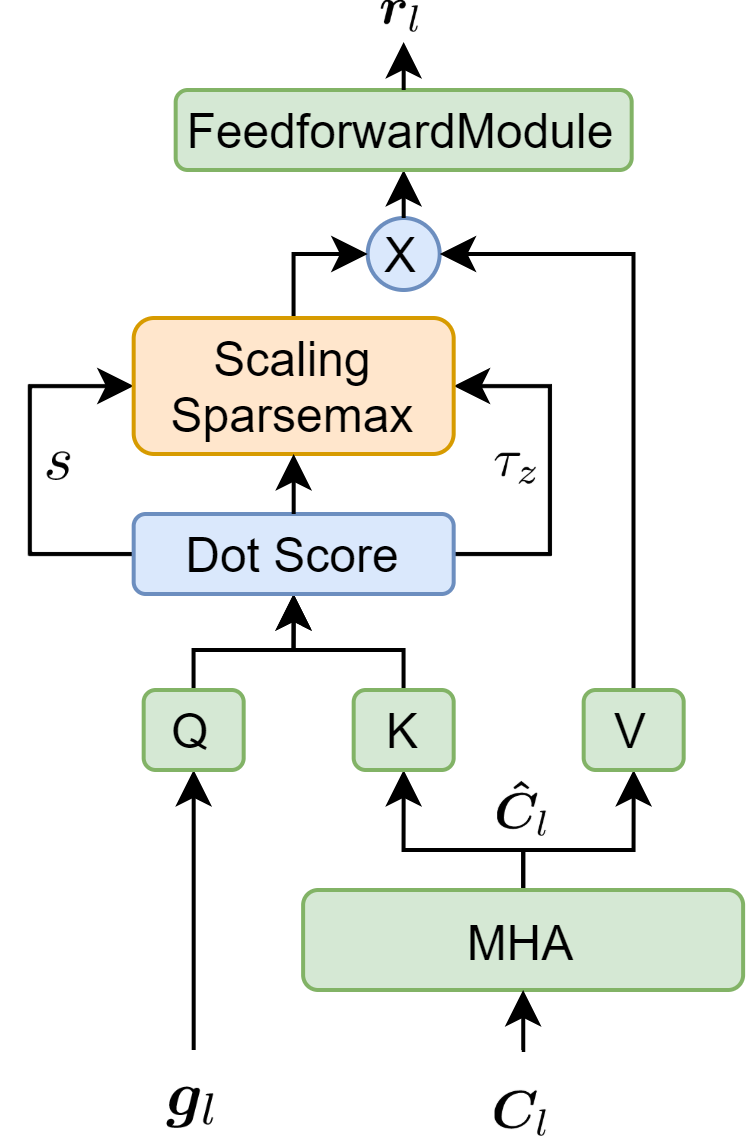}
		\label{fig:scalingsparsemax}
	\end{minipage}%		
	}
	\caption{The structure of three stream attention architectures. (a) Softmax. (b) Sparsemax. (c) Scaling Sparsemax.}
	\label{fig:fuison}
\end{figure}

Fig. \ref{fig:softmax} shows the architecture of the Softmax-based stream attention that has been used in \cite{li2020practical}.

\subsection{Stream attention with Sparsemax} \label{sec: sparsemax}
The common Softmax has a limitation for ad-hoc microphone arrays that its output ${\rm Softmax}_i(z) \neq 0$ for any $z$ and $i$, which can not be used for channel selection. To address this problem, we propose Sparsemax stream attention as shown in Fig. \ref{fig:sparsemax}, where the Sparsemax \cite{martins2016softmax} is defined as:
\begin{equation}
{\rm Sparsemax}(\bm{z}) = \mathop{\arg\min}_{\bm{p}\in \Delta^{K-1}} \Vert \bm{p} - \bm{z} \Vert^2
\end{equation}
where $\Delta^{K-1} = \left\{\bm{p} \in \mathbb{R}^K \ | \sum_{i=1}^{K} p_i = 1, p_i \ge 0 \right\}$ represents a {$(K-1)$-dimensional simplex}. Sparsemax will return the Euclidean projection of the input vector $\bm{z}$ onto the simplex, which is a sparse vector. Its solution has the following closed-form:
\begin{equation}
{\rm Sparsemax}_i(\bm{z}) = \max \left( z_i - \tau(\bm{z}), 0\right)
\end{equation}
where $\tau : \mathbb{R}^K \rightarrow \mathbb{R} $ is a function to find a soft threshold, which will be described in detail in the next section.

\subsection{Stream attention with Scaling Sparsemax}\label{sec: scaling sparsemax}
 From section \ref{sec: sparsemax}, we see that the output of Sparsemax is related to the input vector and the dimension of the simplex. However, in our task, the values of the input vector {vary} in a large range caused by the random locations of the microphones. The dimension of the simplex {is related to} the number of the channels which is also a variable. Therefore, Sparsemax may not generalize well in some cases.

To address this problem, we propose Scaling Sparsemax as shown in Fig. \ref{fig:scalingsparsemax}. It rescales Sparsemax by a trainable scaling factor $s$ which is obtained by:
\begin{align}
s = 1 + {\rm ReLU}({\rm Linear}([\Vert \bm{z} \Vert, C]^T))
\end{align}
where $\Vert \bm{z} \Vert$ is the L2 norm of the input vector, and $\rm{Linear}(\cdot)$ is a $1\times 2$-dimensional learnable linear transform.

Algorithm 1 describes the Scaling Sparsemax operator. When $s=1$, Scaling Sparsemax becomes equivalent to Sparsemax.

\begin{algorithm}[ht]
	\caption{Scaling Sparsemax}
	\label{alg1}
	% \BlankLine
	\KwIn{$\bm{z}$, $s$}
	Sort $\bm{z}$ as $z_{(1)} \ge \cdots \ge z_{(K)}$ \\
	Initialize $k \leftarrow K$ \\
	\While {$k > 0$}
	{
		\If {$z_{(k)} \ge (\sum_{i=1}^{k}z_{(i)} - s) / k$}
		{
			$\tau(\bm{z}) := (\sum_{i=1}^{k}z_{(i)} - s) / k$ \\
			Break \\
		}		
		$k \leftarrow k - 1$ \\
	}
	\KwOut{$\bm{p} \mbox{ where } p_i = \max(z_i - \tau(\bm{z}),0) / s$}
\end{algorithm}

\section{Experiments} \label{sec: exp}

\subsection{Experimental setup}

Our experiments use three data sets, which are the Librispeech ASR corpus \cite{panayotov2015librispeech}, Librispeech simulated with ad-hoc microphone arrays (Libri-adhoc-simu), and Librispeech played back in real-world scenarios with 40 distributed microphone receivers (Libri-adhoc40) \cite{guan2021libriadhoc40}.  Each node of the ad-hoc microphone arrays of Libri-adhoc-simu and Libri-adhoc40 has only one microphone. Therefore, a channel refers to a node in the remaining of the paper. Librispeech contains more than $1000$ hours of read English speech from
{$2484$} speakers. %more than $2000$ speakers.
In our experiments, we selected $960$ hours of data to train single channel ASR systems, and selected $10$ hours of data for development.

Libri-adhoc-simu uses 100 hours `train-clean-100' subset of the Librispeech data as the training data. It uses `dev-clean' and `dev-other' subsets as development data, which contain 10 hours of data in total. It takes 'test-clean' and 'test-other' subsets as two separate test sets, which contain 5 hours of test data respectively. For each utterance, we simulated a room.
The length and width of the room were selected randomly from a range of $[5, 25]$ meters. The height was selected randomly from $[2.7, 4]$ meters. Multiple microphones and one speaker source were placed randomly in the room. We constrained the distance between the source and the walls to be greater than $0.2$ meters, and the distance between the source and the microphones to be at least $0.3$ meters.
We used an image-source model\footnote{https://github.com/ehabets/RIR-Generator} to simulate a reverberant environment and selected T60 from a range of $[0.2, 0.4]$ second.
A diffuse noise generator\footnote{https://github.com/ehabets/ANF-Generator} was used to simulate uncorrelated diffuse noise. The noise source for training and development is a large-scale noise library containing over $20000$ noise segments \cite{tan2020speech}, and the noise source for test is the noise segments from CHiME-3 dataset \cite{barker2015third} and NOISEX-92 corpus \cite{varga1993assessment}.
{We randomly generated $16$ channels for training and development, and $16$ and $30$ channels respectively for test.}

\begin{figure*}[ht]
	\centering
	\includegraphics[scale=0.38]{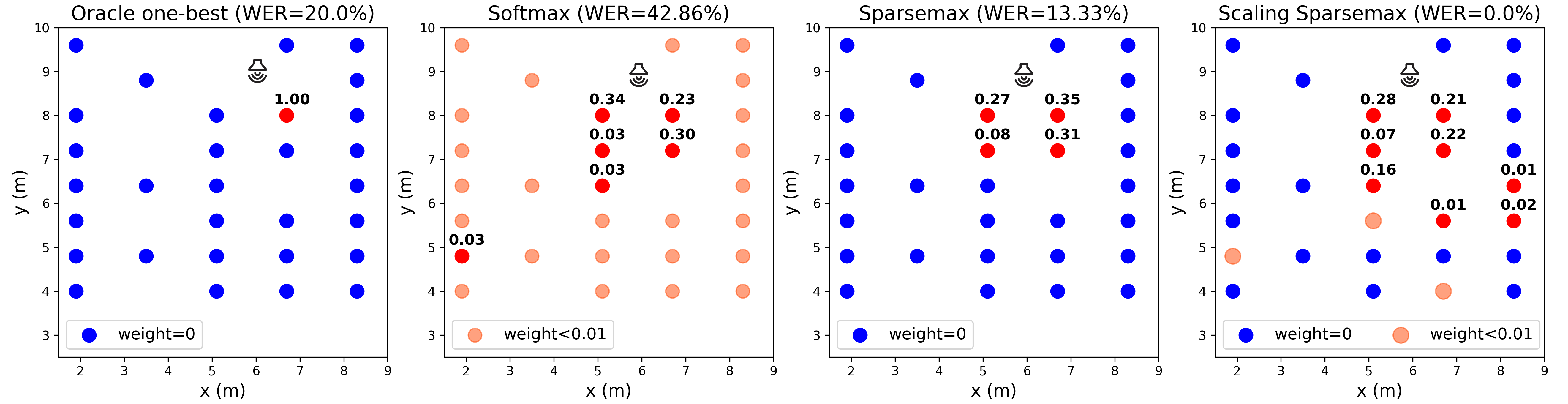}
	\caption{{Visualization of the channel selection results on the utterance ID '3570-5694-0013' of Libri-adhoc40, where each dot represents a microphone, and the number aligned with the dot represents the average weight of the channel over time.}}
	\label{fig:visua}
\end{figure*}

Libri-adhoc40 was collected by playing back the `train-clean-100', `dev-clean', and `test-clean' corpora of Librispeech in a large room \cite{guan2021libriadhoc40}. The recording environment is a real office room with one loudspeaker and $40$ microphones. It has strong reverberation with little additive noise. The positions of the loudspeaker and microphones are different in the training and test set, where the loudspeaker was placed in 9, 4, and 4 positions in the training, development, and test sets respectively.
The distances between the loudspeaker and microphones are in a range of $[0.8, 7.4]$ meters. We randomly selected $20$ channels for each training and development utterances, and $20$ and $30$ channels for each test utterance which corresponds to two test scenarios.

The feature and model structure are described in Table \ref{tab: modeldesc}. In the training phase, we first trained the single channel conformer-based ASR model with the clean Librispeech data. When the model was trained, the parameters were fixed and sent to the multichannel conformer-based ASR model. Finally, we trained the stream attention with the multichannel noisy data. In the testing phase, we used greedy decoding without language model. WER was used as the evaluation metric.

We compared the proposed Sparsemax and Scaling Sparsemax with the Softmax stream attention. Moreover, we constructed an \textit{oracle one-best} baseline, which picks the channel that is physically closest to the sound source as the input of the single channel conformer-based ASR model. Note that the keyword ``oracle'' means that the distances between the speaker and the microphones are known beforehand.

\begin{table}[t]
	\caption{{Descriptions of the acoustic feature and the model structure.}}
	\label{tab: modeldesc}
	\centering
	% \begin{tabular}{ r@{}l  r }
	\scalebox{0.85}{
		\begin{tabular}{lllll}
			\toprule
			\multicolumn{1}{l|}{\multirow{2}{*}{\textbf{Feature}}} &
			\multicolumn{1}{l}{Type: fbank} & \multicolumn{2}{l}{\# dimensions ($D_x$): $80$} & \\ \cline{2-5}
			\multicolumn{1}{l|}{} & \multicolumn{4}{l}{Data augmentation: SpecAugment\cite{park2019specaugment}}\\
			\toprule
			\multicolumn{1}{l|}{\multirow{2}{*}{\textbf{Conformer structure}}} &
			\multicolumn{4}{l}{\# number of blocks: $N_1=12, \ N_2=6$} \\ \cline{2-5}
			\multicolumn{1}{l|}{} & \multicolumn{4}{l}{Vocabulary size ($D_v$): $5000$} \\
			\toprule
			\multicolumn{1}{l|}{\textbf{Multi-head attention}} &
			\multicolumn{1}{l}{\# heads: 8} & \multicolumn{2}{l}{\# dimensions ($D_h$): 512} & \\ % \cline{2-5}
			\toprule
			\multicolumn{1}{l|}{\textbf{Stream attention}} &
			\multicolumn{1}{l}{\# heads: 1} & \multicolumn{2}{l}{\# dimensions ($D_h$): 512} & \\ % \cline{2-5} 		
			\bottomrule
		\end{tabular}
	}	
\end{table}

%\begin{table}[t]
%	\caption{Descriptions of the acoustic feature and the model structure.}
%	\label{tab: modeldesc}
%	\centering
%	% \begin{tabular}{ r@{}l  r }
%	\scalebox{0.9}{
%		\begin{tabular}{llll}
%			\toprule
%			\multicolumn{4}{c}{\textbf{Feature}} \\ \hline
%			Type & \# dimensions & \multicolumn{2}{l}{Data augmentation} \\ \hline
%			fbank & $D_x=80$ & \multicolumn{2}{l}{SpecAugment\cite{park2019specaugment}} \\
%			\toprule
%			\multicolumn{4}{c}{\textbf{Conformer structure}} \\ \hline
%			\multicolumn{2}{l}{\# blocks} & \multicolumn{2}{l}{Vocabulary size} \\ \hline
%			\multicolumn{2}{l}{$N_1=12$,\ $N_2=6$} & \multicolumn{2}{l}{$D_v=5000$} \\
%			\toprule
%			\multicolumn{2}{c|}{\textbf{MultiHeadAttention}} &
%			\multicolumn{2}{c}{\textbf{StreamAttention}}\\ \hline
%			\# heads & \multicolumn{1}{l|}{\# dimensions} & \# heads & \# dimensions \\ \hline
%			$n=8$ & \multicolumn{1}{l|}{$D_h=512$} & $n=1$ & $D_h=512$ \\		
%			\bottomrule
%		\end{tabular}
%	}	
%\end{table}

\subsection{Results} \label{sec: result}

% \subsubsection{Evaluation with Libri-adhoc-simu}
\begin{table}[t]
	\caption{Comparison results on \textbf{Libri-adhoc-simu} (in WER (\%)). The term ``ch'' is short for channels in test.}
	\label{tab: simu}
	\centering
	% \begin{tabular}{ r@{}l  r }
\scalebox{0.9}{
	\begin{tabular}{lcccc}
		\toprule
		% \multicolumn{5}{l}{\bf{Feature}} \\
		\multirow{2}{*}{Method}  & \multicolumn{2}{c}{test-clean} & \multicolumn{2}{c}{test-other} \\
		& 16-ch & 30-ch & 16-ch & 30-ch \\
		\midrule
		Oracle one-best & $14.3$ & $10.6$ & $30.1$ & $24.5$ \\
		Softmax & $15.4$ & $11.8$ & $33.7$ & $28.9$ \\
		Sparsemax (proposed) & $11.5$ & $8.3$ & $27.5$ & $22.9$ \\
		ScalingSparsemax (proposed) & $\mathbf{10.7}$ & $\mathbf{7.8}$ & $\mathbf{26.5}$ & $\mathbf{21.4}$ \\
		\bottomrule
	\end{tabular}
	}
\end{table}

Table \ref{tab: simu} lists the performance of the comparison methods on Libri-adhoc-simu. From the table, we see that (i) all three stream attention methods perform good in both test scenarios. Particularly, the generalization performance in the mismatched $30$-channel test environment is even better than the performance in the matched $16$-channel environment. It also demonstrates the advantage of adding channels to ad-hoc microphone arrays. (ii) Both Sparsemax and Scaling Sparsemax achieves significant performance improvement over Softmax. For example, Scaling Sparsemax stream attention achieves a relative WER reduction of $33.90$\% over Softmax on the `test-clean' set and $26.0$\% on the `test-other' set of the 30-channel test scenario.
% \subsubsection{Evaluation with Libri-Real}
\begin{table}[t]
	\caption{Comparison results on the \textbf{Libri-adhoc40} semi-real data (in WER (\%)).}
	\label{tab: real}
	\centering
	% \begin{tabular}{ r@{}l  r }
\scalebox{0.9}{
	\begin{tabular}{lcc}
		\toprule
		% \multicolumn{5}{l}{\bf{Feature}} \\
		Method & 20-ch & 30-ch \\	
		\midrule
		Oracle one-best & $28.2$ & $22.5$  \\
		Softmax & $29.7$ & $25.4$ \\
		Sparsemax (proposed)& $33.7$ & $30.3$ \\
		ScalingSparsemax (proposed)& $\mathbf{23.3}$ & $\mathbf{19.3}$ \\
		\bottomrule
	\end{tabular}
}
\end{table}

Table \ref{tab: real} shows the results on the Libri-adhoc40 semi-real data. From the table, one can see that the proposed Scaling Sparsemax performs well. It achieves a relative WER reduction of $17.4$\% over the ``oracle one best baseline'' on the 20-channel test scenario, and $14.2$\% on the mismatched 30-channel test scenario.

%\begin{figure}[t]
%	\centering
%	\includegraphics[scale=0.3]{figure/all_visu.png}
%	\label{fig:v_s_sp}	
%	\caption{Visualization of the channel selection results on the utterance ID '3570-5694-0013' of Libri-adhoc40, where each dot represents a microphone and the number aligned with the dot represents the average weight of the channel over time.}
%	\label{fig:visua}
%\end{figure}
Fig. \ref{fig:visua} shows a visualization of the channel selection effect on an utterance of Libri-adhoc40. From the figure, we see that (i) Softmax only considers channel reweighting without channel selection; (ii) Although Sparsemax conducts channel selection, its channel selection method punishes the weights of the channels too heavy. (iii) Scaling Sparsemax only sets the weights of very noisy channels to zero, which results in the best performance.

\section{Conclusions} \label{sec: conclu}
In this paper, we propose two channel selection methods in the conformer-based ASR system for optimizing the ASR performance with ad-hoc microphone arrays directly. Specifically, we replace the Softmax operator in the stream attention with Sparsemax to make it capable of channel selection. Because Sparsemax punishes the weights of channels severely, we propose Scaling Sparsemax to punish the weights mildly, which only sets the weights of very noisy channel to zero. We evaluate our model on a simulation data set with background noise and a semi-real data set with high reverberation. Experimental results show that the proposed Scaling Sparsemax stream attention not only outperforms the Softmax steam attention but also the oracle one-best, in both simulated data and a semi-real corpus. The results also demonstrate the importance of channel selection to speech recognition with large-scale ad-hoc microphone arrays.
% \section{Acknowledgements}

\newpage

\bibliographystyle{IEEEtran}

\bibliography{mybib}

% \begin{thebibliography}{9}
% \bibitem[1]{Davis80-COP}
%   S.\ B.\ Davis and P.\ Mermelstein,
%   ``Comparison of parametric representation for monosyllabic word recognition in continuously spoken sentences,''
%   \textit{IEEE Transactions on Acoustics, Speech and Signal Processing}, vol.~28, no.~4, pp.~357--366, 1980.
% \bibitem[2]{Rabiner89-ATO}
%   L.\ R.\ Rabiner,
%   ``A tutorial on hidden Markov models and selected applications in speech recognition,''
%   \textit{Proceedings of the IEEE}, vol.~77, no.~2, pp.~257-286, 1989.
% \bibitem[3]{Hastie09-TEO}
%   T.\ Hastie, R.\ Tibshirani, and J.\ Friedman,
%   \textit{The Elements of Statistical Learning -- Data Mining, Inference, and Prediction}.
%   New York: Springer, 2009.
% \bibitem[4]{YourName17-XXX}
%   F.\ Lastname1, F.\ Lastname2, and F.\ Lastname3,
%   ``Title of your INTERSPEECH 2021 publication,''
%   in \textit{Interspeech 2021 -- 20\textsuperscript{th} Annual Conference of the International Speech Communication Association, September 15-19, Graz, Austria, Proceedings, Proceedings}, 2020, pp.~100--104.
% \end{thebibliography}

\end{document}